\documentclass[amsmath,amssymb,prb,twocolumn]{revtex4}
\usepackage{graphicx}
\usepackage{float}
\usepackage{color}
\usepackage{lipsum}
\begin{document}
\title{Metamorphoses of the flow past an obstacle of  a resonantly-driven bistable polariton fluid }
\author{Vincent Hakim}
\email{vincent.hakim@ens.fr}
\affiliation{Laboratoire de Physique de l'Ecole Normale Sup\'erieure, CNRS,  Ecole Normale Sup\'erieure, PSL University, Sorbonne Universit\'e,
Universit\'e Paris-Diderot, Paris, France}
\author{Simon Pigeon}
\email{simon.pigeon@lkb.upmc.fr}
\affiliation{Laboratoire Kastler Brossel, Sorbonne Universit\'e, CNRS,  Ecole Normale Sup\'erieure, PSL University, Collège de France, Paris, France}
\author{Amandine Aftalion}
\email{amandine.aftalion@ehess.fr}
\affiliation{Ecole des Hautes Etudes en Sciences Sociales,  Centre d'Analyse et de Math\'ematique Sociales, UMR-8557, Paris, France.}

\date{\today}
\begin{abstract}
Motivated by recent experiments, we theoretically analyze  the flow past an obstacle of a one-dimensional "quantum fluid of light" which is resonantly driven, and exhibits bistability. The flow is found to abruptly change several
 times when the fluid velocity or the obstacle potential strength is increased. These transitions display  unusual features. In contrast to the cases of usual fluids and superfluids, the transitions take place between stationary states. They involve the fluid bistability in an essential way. Remarkably, at the transitions points,  the fluid in the obstacle wake lies in the unstable intermediate density  state.
\end{abstract}

\maketitle
\section{Introduction}
The discovery of Bose-Einstein condensation has opened a very active field of research
\cite{pitaevskii2016}. Besides cold atomic vapors,
Bose-Einstein condensation has also been achieved in exciton-polariton fluids  \cite{kasprzak2006bose,amelio2020}.
These ``quantum fluids of light'' \cite{carusotto2013} result from the strong coupling between the excitonic resonance of a semiconductor quantum well and a microcavity  electromagnetic field.
Their solid-state nature and  the higher condensation temperature associated with the polariton very low mass turn  them into attractive systems.
In early  experiments, polaritons were created with  a transient \cite{Nardin2011} or  a spatially localized \cite{amo2011polariton}
driving field to avoid interfering with the superfluid behavior of the condensate. The short polariton lifetime
then restricted the experiment duration or limited the observations to a local region around the pumping spot. In order to bypass these limitations, it has been found useful
to introduce a weaker  resonant drive, a so-called ``support field'',  away from the strong localized pumping spot used to create the polaritons \cite{pigeon2017}.  The extended quasi-resonant drive tends to lock the phase of the condensate  and
its dynamics, which is different from the dynamics of a conventional fluid or of a superfluid \cite{juggins2018}.
When  the support field is not too strong, it allows the formation  of collective excitations, such as vortices \cite{koniakhin2019} and  dark solitons \cite{maitre2020dark}.
This new coherently driven
regime has started to be investigated theoretically \cite{pigeon2017, chestnov2019,amelio2020,pa2021,joly2021} and experimentally \cite{stepanov2019, koniakhin2019,lerario2020vortex,maitre2020dark}.

Here, we consider the flow of a resonantly-driven condensate past an obstacle. Such a set-up has been the subject of many investigations both for classical fluids  \cite{williamson1996} and superfluids \cite{pitaevskii2016}. In two or three dimensions, the flow becomes unsteady at a critical velocity through an oscillatory (Hopf) bifurcation. For superfluids, this leads to the nucleation of vortices \cite{raman99} past the critical velocity. In a one-dimensional  setting, gray solitons are generated and propagate from the obstacle along the flow \cite{engels2007}. For standard (conservative) superfluids, these dynamical behaviors are well-described in the framework of the Gross-Pitaevskii equation \cite{fpr1992, vh1997, JMA, HB, ADP}. If the creation of defects in the flow of a resonantly-driven condensate has been observed, strong deviation from standard superfluids behaviors were reported \cite{pigeon2017,amelio2020,pa2021, lerario2020vortex}. It remains to better understand the transition and its dependence on the fluid bistability \cite{baas2004}, and, more generally, condensate dynamics in the presence of resonant drive and dissipation.

In the present work, we focus on the one-dimensional case which is easier to analyze than higher dimensional cases. We find that multiple transitions in the flow occur  when the fluid velocity is increased, or when the obstacle strength is increased at fixed velocity.
We show that these transitions are of a very different type from the usual ones in fluids and superfluids. Moreover,
 their unusual character
 forbids their prediction
 from the characteristics of excitations around the steady flow, as
 done for superfluids with the Landau criterion \cite{pitaevskii2016}.

\section{The generalized Gross-Pitaevskii equation and bistability}
We consider the
fluid  described by the following generalized Gross-Pitaevskii equation (GGPE)
\begin{equation}
i \hbar \partial_t \psi=\frac{-\hbar^2}{2m}\partial^2_{x} \psi +\left[V(x)- \hbar\Delta -i \frac{\hbar\gamma}{2} + g \vert \psi \vert^2\right] \psi + F e^{i k_p x}.
\label{eqdef}
\end{equation}
In the context of exciton-polariton microcavity physics, Eq.~(\ref{eqdef}) provides an effective description of a driven lower polariton field \cite{carusotto2013,amelio2020}, with the polariton-polariton repulsive interaction accounted by the constant $g>0$. Additional terms as compared to the usual GPE arise from the coherent drive and dissipation \cite{carusotto2013,amelio2020}.  The support field
is characterized by its
 amplitude $F$, its momentum $k_p$, produced by a slight tilt of the driving laser beam with respect to the cavity plane and the detuning $\Delta$ of its frequency from the lower polariton band bottom frequency.  Dissipation is described by the rate $\gamma>0$ arising from the polariton finite lifetime.
 The potential $V(x)$ models a localized  obstacle. It is our main aim to characterize its effect on the fluid flow described by Eq.~(\ref{eqdef}). It is worth noting that our results are also relevant for nonlinear optics \cite{pomeau1993} where  Eq.~(\ref{eqdef})
 is  known as the Lugiato-Lefever equation \cite{lugiato1987} and describes the wave evolution  in a cavity filled with a nonlinear medium (see e.g. \cite{parra2016dark} and ref. therein).

The explicit $x$-dependence in Eq.~(\ref{eqdef}) can be eliminated by defining,
\begin{equation}
\psi= \sqrt{\hbar\gamma/2g}\, \phi(x) \exp(i k_p x).
\end{equation}
The function $\phi$ then obeys the equation,
\begin{equation}
i \partial_{\tau} \phi= -\frac{1}{2}\partial_{yy} \phi -i k_0 \partial_y \phi\\ -\left[\delta_V(y)+i - \vert \phi \vert^2\right] \phi + f ,
\label{eq2}
\end{equation}
where we have introduced the dimensionless variables
$y=x \sqrt{m\gamma/2\hbar}, \ \tau= t \gamma/2$, and  constants, $k_0= k_p \sqrt{2\hbar/m\gamma}, f= F  \sqrt{g}(2/\hbar\gamma)^{3/2}$ , and defined  the function
\begin{equation}
\delta_V(y)= \delta_0- \frac{2}{\hbar\gamma} V(y),\ \mathrm{with} \  \delta_0=\frac{2}{\hbar\gamma} \left[\hbar \Delta-\frac{(\hbar k_p)^2}{2m}\right].
\label{bdef}
\end{equation}

Before considering the effect of a localized obstacle, we briefly recall some properties of the fluid described by Eq.~(\ref{eq2}).
When $V(y)=0$ and $\delta_V(y)=\delta_0$, Eq.~(\ref{eq2}) has constant solutions in space and time with a homogeneous density $\rho=  \vert \phi \vert^2$ which can readily  be seen to simply obey,
 \begin{equation}
B(\rho):= [(\rho -\delta_0)^2 +1] \rho =f^2 .
\label{fpe1}
\end{equation}

Two cases can be distinguished. When $\delta_0<0$, the function $B(\rho)$, defined in Eq.~(\ref{fpe1}), is increasing from $0$ to $+\infty$
 with the density.
 As a consequence, the density $\rho$ is also an increasing function of the forcing amplitude $f$.
 When $\delta_0>0$, namely for blue detuning,
 $B(\rho)$ can be non-monotonic with multiple homogeneous solutions for a given forcing. A simple analysis of Eq.~(\ref{fpe1}) shows that this actually happens when
 $\delta_0>\sqrt{3}$. An example of this S-like dependency of the density with the driving field is plotted in Fig.~\ref{fullsim}a.
  In this case, three solutions exist in a window of intermediate forcing strengths i.e. for $ B(\rho_+)<f^2 < B(\rho_-)$ with
 \begin{equation}
\rho _{\pm}= \frac{1}{3} \left(2 \delta_0 \pm \sqrt{\delta_0^2 -3}\right) .
\label{d0vsrho}
\end{equation}
The high density (HD) and low density (LD) solutions are stable while the intermediate density (ID) one is unstable, as explicitly shown in Appendix \ref{appA}.
Bistability stems from the  positive feedback between the fluid density increase and forcing efficiency, for blue detuning. When the density of the fluid increases, 
the detuning of the forcing decreases as a consequence  of the repulsive self-interactions, as can explicitly be seen in Eq.~(\ref{fpe1}). This results in a more closely resonant and thus more efficient forcing which, in turn,  increases the fluid density.
 This bistability for sufficiently strong blue detuning is well-known in nonlinear optics \cite{carusotto2013} and has been demonstrated for polaritons in microcavities
 \cite{baas2004}.
 While in the  LD state, self-interactions are unimportant,  the strong self-interactions in the HD state  modify the fluid  flow properties
 \cite{amo2009superfluidity,juggins2018}.
\section{Flow past an obstacle : numerical simulations and flow metamorphosis}
\label{numsim}
\begin{figure}[htbp]
\begin{center}
\includegraphics[height=1.\textwidth]{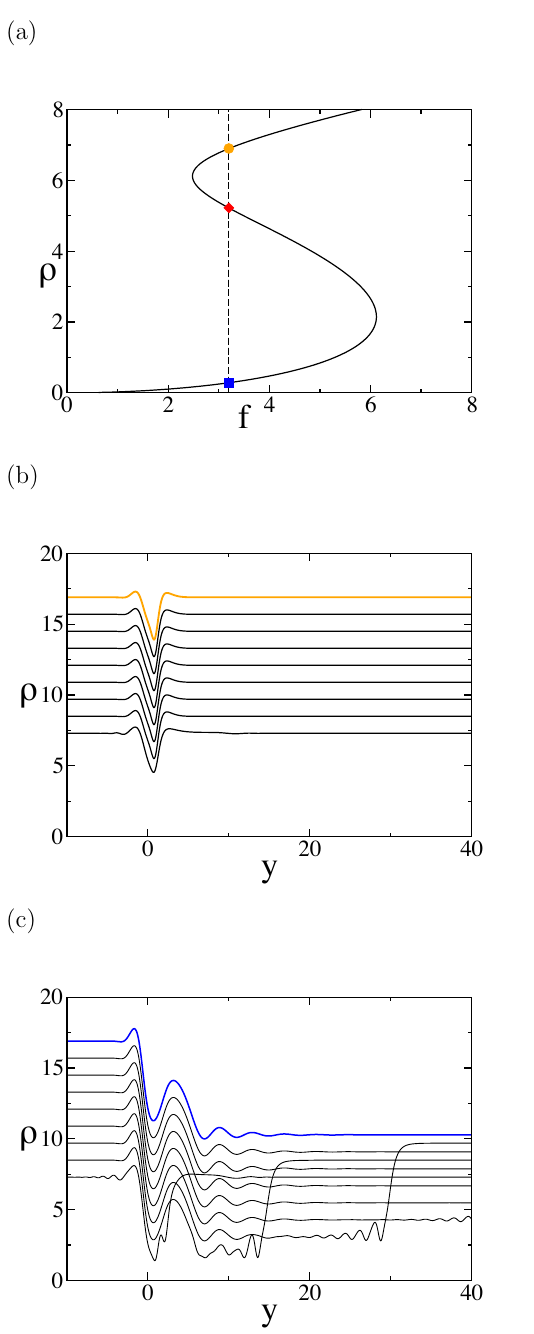}
\caption{Numerical simulations of Eq.~(\ref{eq2}). (a)  Fluid density $\rho$ vs. forcing $f$ as described by Eq.~(\ref{fpe1}). In the parameter regime considered, there are three  homogeneous steady states, stable HD  (high density, solid orange circle) and LD states (low density, solid blue square) and an unstable ID one (intermediate density, solid red diamond). (b)\& (c) The fluid is injected in the HD state. The fluid density is shown at successive times separated by  $\Delta\tau=6$. Successive curves are shifted upward with time by 1.2 unit of density to highlight that the flow becomes stationary. The last simulation curve is shown as a thicker colored line. (b)
for a potential amplitude $u_m=2$, the flow is steady with  the wake of the obstacle in the HD state at $y\gg 0$. (c) For a larger $u_m=6$, the flow is still stationary, but is in the LD state in the wake of the obstacle.
Other parameters are $\delta_0=6.2, f=3.2, k_0=2.75$ which corresponds to the
 typical experimental values $\hbar \Delta=0.5 meV, \hbar \gamma=0.1 meV, \hbar^2/m=1meV \mu m^2, k_p=0,616 \mu m^{-1}, \sqrt{g}F=0.036 (meV)^{3/2}$. The potential range is
 $\sigma=1$ corresponding to $4.5 \mu m$, the chosen unit length.
}
\label{fullsim}
\end{center}
\end{figure}

 Having recalled  the basic features of the homogeneous state, we proceed and describe our simulations of Eq.~(\ref{eq2}) with a localized repulsive ($V(y)>0$) Gaussian potential
 \begin{equation}
V(y) = \frac{\hbar \gamma}{2} u_m  \exp[ -(y/\sigma)^2] .
\label{gpot}
\end{equation}
 We focus on the bistable parameter regime with  $\delta_0>\sqrt{3}$ and the forcing $f$ in the appropriate intermediate interval (see Fig.~\ref{fullsim}a).
 In an experimental setting,  a strong driving field in a far upstream local region would be used to create the HD state as proposed in ref. \cite{pigeon2017}, experimentally realized in e.g. \cite{lerario2020vortex}, and sketched in Fig.~\ref{fS0}. Instead, here, we study an equivalent but mathematically simpler situation by simply setting up the fluid in the HD state as an upstream boundary condition.

 Simulations of Eq.~(\ref{eq2}) with the Gaussian potential (\ref{gpot}) are performed as in ref. \cite{vh1997} with a finite-difference semi-implicit Crank-Nicholson scheme. The reported results are obtained in a symmetric domain around the origin,  of linear size 150 or 200, with a spatial step $\Delta y=0.05$ and a time step $\Delta \tau= 10^{-4}$.

 For a low potential amplitude, the flow is steady. The density decreases as expected in the region of the repulsive potential, and it returns smoothly to the HD state in the wake of the obstacle, as shown in Fig.~\ref{fullsim}b. For a weak potential, this configuration has been previously studied perturbatively in the context of a moving particle in a polariton fluid \cite{vanregemortel2014,minguzzi2020}.  An increase in the potential amplitude $u_m$ produces a transition in the flow, as shown
 in Fig.~\ref{fullsim}c.  However, the character of the transition appears to be very different from the usual transitions to time-dependent flows in fluids and superfluids.
 Instead, above the transition the flow is still stationary, after a transient, but with the fluid density in the LD state downstream of the obstacle, as shown Fig.~\ref{fullsim}c. In other words, for the driven-dissipative GGPE, the steady flow undergoes
  a metamorphosis instead of becoming time-dependent.
 That the flow density downstream of the obstacle lies in the LD  state provides a first hint that the fluid bistability is playing a significant role in the observed transition. 

\begin{figure*}
\begin{center}
\includegraphics[width=.70\textwidth]{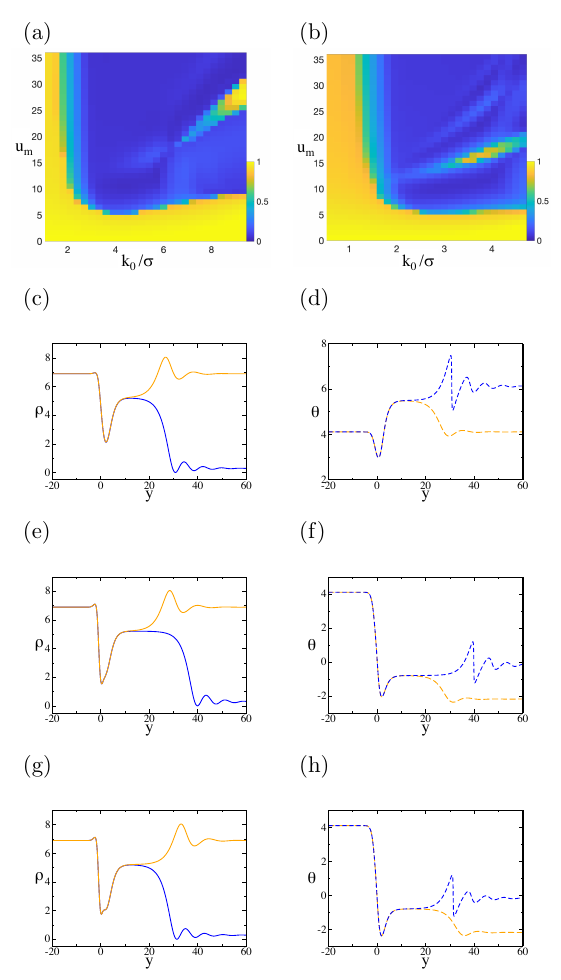}
\caption{Results of simulations of the GGPE (Eq.~(\ref{eqdef}), (\ref{eq2}))  for  $f=3.2, \delta_0=6.2$ for different values of the localized potential amplitude $u_m$ and velocity $k_0$.
The fluid is in the HD-state at $y\ll0$. (a)  The potential range
 is $\sigma=1$ in (a) and $\sigma=2$ in (b). The  color code indicates the relative density in the wake of the obstacle relative to the density in the HD-state. Intermediate colors are due to the limited resolution of the numerical procedure used to scan this two-parameter plot.
 (c-h) For $\sigma=2$, close-up with higher resolution, of 3 transitions that take place when  $u_m$ is increased on the vertical line $k_0=6.4$ of panel (b). 
 Solution densities  ((c),(e),(g), solid lines) and phases ((d),(f),(h), dashed lines) are shown
for two values of the potential just below and just above the transition. The fluid  in the obstacle wake is either in the HD-state (orange) of the LD-state (blue).
(c) \& (d)
 1st transition with $u_m=4.85$ (HD state) and $u_m=4.86$ (LD state)
(e) \&(f) 2nd transition  with $u_m=14.16$ (LD state) and $u_m=14.17$ (HD state).
The 2nd transition is inverted as compared to the 1st, namely the fluid density jumps back to the HD state when $u_m$ is increased. (g) \&(h) 3rd transition  with $u_m=14.84$ (HD state) and  $u_m=14.85$ (LD state).  The transition is analogous to the first one, but with the phases shifted by $2\pi$ in the far downstream wake of the obstacle.
 }
\label{fS3}
\end{center}
\end{figure*}

Simulations of Eq.~(\ref{eq2})  for different potential amplitudes $u_m$ and different flow velocities $k_0$,  provide a more global view of the dynamical regimes of the  GGPE flow past an obstacle, as summarized in Fig.~\ref{fS3}. The results are displayed for two values of the potential range $\sigma=1$   (Fig.~\ref{fS3}a) and $\sigma=2$   (Fig.~\ref{fS3}b)
\footnote{The results of numerical simulations of the GGPE (Eq.~(\ref{eq2}))  were obtained by using the semi-implicit Crank-Nicholson code described above, in all figures except Fig.~\ref{fS3}a,b. Fig.~\ref{fS3}a,b  were produced by using a  Runge-Kutta code with periodic boundary conditions in a box of size $L=38.6823$ with a spatial step  $\Delta y= 0.0377$ and a time step $\Delta t = 1.8665 \ 10^{-05}$. The high density state was created by a localized zone of strong forcing upstream of the obstacle as in ref. \cite{pigeon2017, pa2021}  and sketched in Fig.~S1. 
Exploratory and additional simulations were also performed with this other code.}
They are qualitatively very similar in the two cases. As expected, the transition point described above extends to a full boundary delimiting two domains  in the $(k_0, u_m)$ plane,  with different kinds of steady state flow. In the outside domain (yellow domain in  Fig.~\ref{fS3}a,b), flows starting in the HD-state upstream of the obstacle return to the HD-state in the wake of the obstacle. On the contrary, in the inside domain (blue domain in  Fig.~\ref{fS3}a,b), flows starting in the HD-state end up in the LD-state in the wake of the obstacle. However, the survey of an extended part of the  $(k_0, u_m)$ plane brings a surprise :  other transitions are found, corresponding in  Fig.~\ref{fS3}a,b to the boundaries of the smaller yellow regions  inside the blue domain. In these smaller regions,
the fluid in the wake of the obstacle is again in the HD-state. These multiple transitions are illustrated in Fig.~\ref{fS3}c-h for $\sigma=2$, by increasing the potential amplitude $u_m$  at fixed flow velocity $k_0$. The fluid density is in  the the HD-state in the wake of the obstacle for low $u_m$ values.  At a first critical value of $u_m$, the fluid in the obstacle wake  jumps in the LD-state  (Fig.~\ref{fS3}c,d)  as described above. When the potential amplitude is further increased a second transition is found, at which the fluid density in the obstacle wake, jumps back to the HD-state (Fig.~\ref{fS3})e,f). At a still higher value of $u_m$, there is a third transition, similar to the first one, where the fluid density in the obstacle wake returns to the LD-state (Fig.~\ref{fS3}e,f).
 
\begin{figure*}
\begin{center}
\includegraphics[width=.85\textwidth]{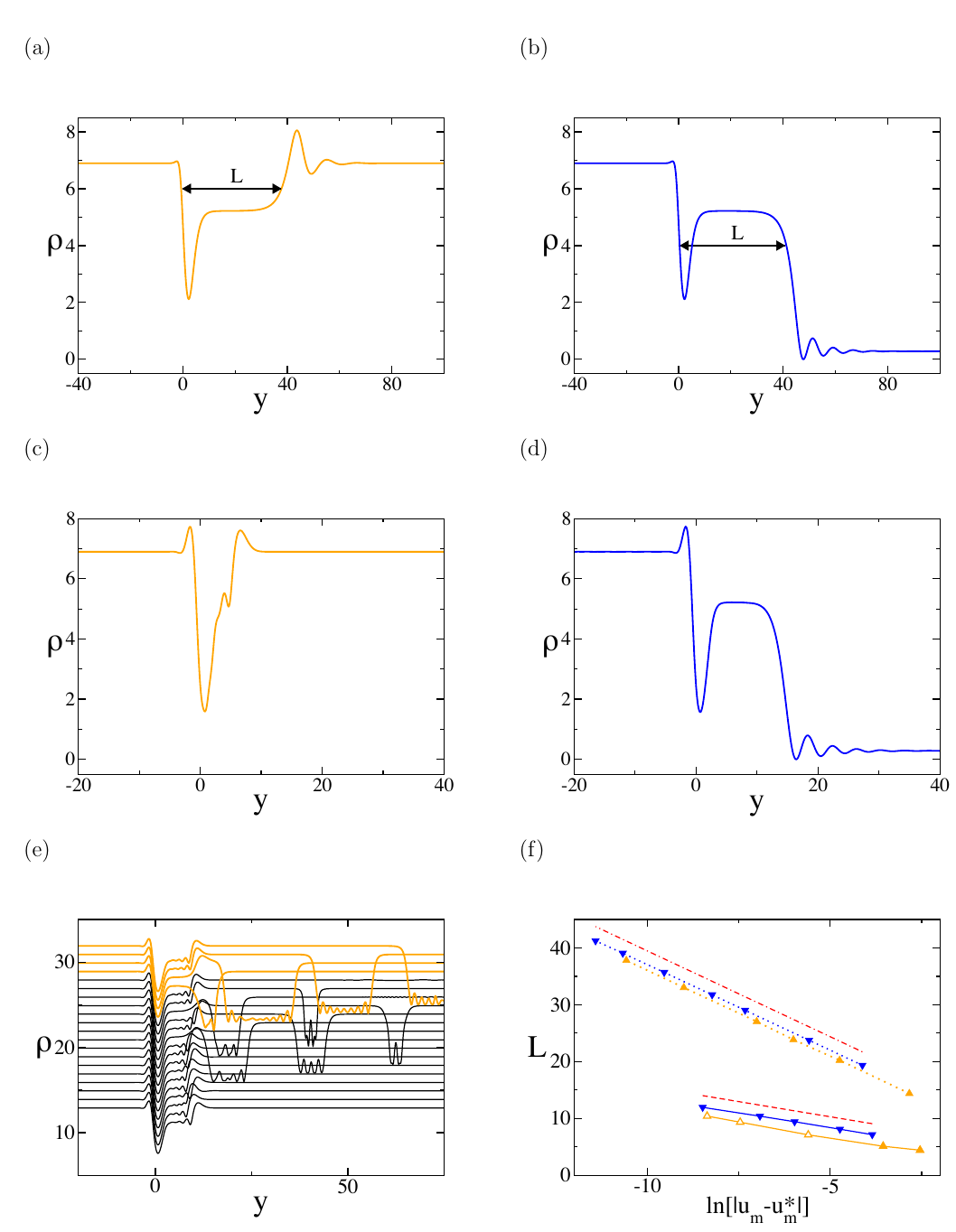}
\caption{Detail of the flow metamorphosis in simulations of Eq.~(\ref{eq2}). (a)\&(b)  $\sigma=2, k_0=6.4$ with (a) $u_m=4.8586914$,  just below the transition with the fluid in the far wake of the obstacle in the HD-state, and (b) $u_m=4.858728$, just above the transition with the fluid in the far wake of the obstacle in the LD-state (see [36])). In both cases, the fluid in the near wake of the obstacle is in the unstable ID-state, as described in the main text. (c), (d) \& (e)
 Same as (a) and (b) for  $\sigma=1$ and three different potential amplitudes: (c) $u_m=5.46$
below the transition, and  (d)
$u_m=5.4789$, just above the transition. 
(e) For $u_m=5.4781$,  below but very close to the critical potential amplitude   $u_m*\simeq 5.4786$, the fluid density is shown at successive times. The curves (dark solid lines) are plotted every 8 time units and shifted upward with time by 1 unit of density. The 4 last curves (thick orange solid lines) are highlighted in order to show the time dependence of the flow. 
  (f) The  length $L$ of the ID state region (depicted in (a) \& (b)) is shown as a function of $\vert u_m-u_m^*\vert$ for   $u_m<u_m^*$ (upward-pointing triangle with orange dotted line ) and $u_m>u_m^*$ (downward-pointing triangle with orange dotted line) when $\sigma=2, k_0=6.4$  and $u_m*\simeq 4.8587$. The predicted asymptotic slope of $-3.02$ (Eq.~(\ref{asid}) and Appendix \ref{appA}) is displayed (dashed-dotted red line).
 The length $L$ is also shown when $\sigma=1, k_0=2.75$ and $u_m^*\simeq 5.4786$ 
  for $u_m<u_m*$ for  
  (upward-pointing triangles with solid orange line) or  $u_m>u_m^*$ (blue solid line and downward-pointing triangle).
  when   $\sigma=1, k_0=2.75$ and $u_m^*\simeq 5.4786$. The predicted asymptotic slope of $-1.07$ 
   is displayed (dashed red line). Filled symbols correspond to steady solutions. Empty symbols corresponds to time-dependent solutions  and are only indicative since the fronts have significant oscillations. 
 In both cases, the length $L$  is defined as the distance from the potential maximum at $y=0$  to the point of density $\rho=6$ (resp. $\rho=4$) of the front joining the ID obstacle wake to the HD (resp. LD) state as shown in panel (a) (resp.~ (b)). In all panels,
 the parameters  $\delta_0=6.2, f=3.2$ are the same as in Fig.~\ref{fullsim}.
}
\label{fscloseup}
\end{center}
\end{figure*}

How does the transition take place in the obstacle wake,  between a steady flow in the HD-state   and a solution in the LD-state, when parameters are varied?
In order to shed light on this question, simulations  very close to a transition point on the lowest transition line are shown
in Fig.~\ref{fscloseup}a-d , for the two potential ranges  $\sigma=1$ and $\sigma=2$. The flow velocity $k_0$ is fixed and the amplitude $u_m$ of the potential is varied.

As shown in Fig.~\ref{fscloseup}a,  for $\sigma=2$, when the amplitude of the potential is  close to, but below, the critical potential amplitude $u_m^*$, the obstacle is followed by a  fluid region of length $L$, close to  the intermediate density (ID) unstable state. This region terminates by a front that joins the ID-state  to the more downstream  HD-state. As the potential amplitude  approaches $u_m^*$, this front stands farther downstream from the obstacle, with an increasing region of the fluid downstream of the obstacle in the unstable ID state.  For potentials with an amplitude slightly greater than $u_m^*$, the complementary process is observed, as shown in
Fig.~\ref{fscloseup}b.
As for subcritical potentials, the obstacle is followed by a fluid region in the unstable ID state but which terminates by a front  joining it  to the stable  LD -state. When increasing the potential amplitude, this front stands closer to the obstacle. It reaches the obstacle and disappears, as soon as the potential amplitude departs by a small amount from $u_m^*$.   These observations strongly suggest that the critical solution is such that the fluid  downward wake exactly stands at the unstable  ID-state.

For $\sigma=1$, the scenario of the transition is similar but with an additional complication. When $u_m$ approaches $u_m^*$ from above, the ID to LD state stands farther and farther downstream from the obstacle with a large region of fluid in the ID state (Fig.~\ref{fscloseup}d) exactly as for $\sigma=2$. As in this previous case, this strongly suggests that the critical solution is such that the fluid density lies in the unstable ID state in the obstacle wake. When $u_m$ is below  $u_m^*$, and approaches it closely, the ID state appears in the obstacle wake together with a front linking it to the HD-state (Fig.~\ref{fscloseup}c). However, the  HD front recedes but becomes non-stationary, when $u_m$  approaches even more closely  $u_m^*$.  This leads to large excursions in density, to and back from the LD-state, that travel in the far wake of the obstacle (Fig.~\ref{fscloseup}e). This phenomenon which only takes place in a very small interval of $u_m$ values below $u_m^*$  is observed for different discretization steps, simulation box sizes and total simulation times. It thus appears real and not due to a numerical instability or to an incomplete relaxation to a steady state.

\section{Critical solutions existence and sharpness of the transitions}
The transitions observed in the numerical simulations suggest that  the  critical flows are  such that, surprisingly, the fluid lies in the unstable ID-state in the far wake of the obstacle. This leads us to consider at which conditions such  steady solutions of Eq.~(\ref{eq2}) that start in the HD-state at $y=–\infty$ and end in the  LD-state at $y=+\infty$ can exist. Another remarkable fact is the sharpness of the observed
transitions when  the potential strength  $u_m$ is varied (see e.g the very small difference between the values of $u_m$ in Fig.~\ref{fscloseup}a and b). We show below that considering the spatially growing modes  around the homogeneous states at $y=-\infty$ and $y=+\infty$  sheds light on both questions.
 
 A stationary solution of Eq.~(\ref{eq2}) obeys a 2nd-order complex equation. Thus, its asymptotic behavior around a homogeneous state  is described by  4  real modes. 
 As shown in Appendix \ref{appA},   three of these four modes are spatially diverging as $y\rightarrow +\infty$ when the stationary solution  is linearized around the  ID-state.
Similarly, there are two diverging modes as $y\rightarrow -\infty$ when the solution is linearized around the HD-state. Let us suppose  integrating in space the time independent version of Eq.~(\ref{eq2}) from the HD-state at $y =-\infty$.  In order for the solution to tend towards the ID-state at $y=+\infty$, the prefactors of the three diverging modes  should be set to zero. However,
the only integration  freedom  lies in the prefactors of the  two convergent modes at $y =-\infty$ since
the amplitudes of the two divergent modes at $y=-\infty$ are  set to zero by the initial condition.  The solution can tend towards the ID-state at $y=+\infty$ only  if  the potential amplitude is used as an additional variable to be adjusted to cancel the three divergent modes. Therefore, a stationary solution linking the HD-state at $y =-\infty$ to the ID-state at $y =+\infty$,  only exists for a discrete set of critical potential amplitudes when other parameters are fixed.  The numerics of Fig.~\ref{fS3} shows that this set actually comprises several values. We also show it analytically in section \ref{asymp.sec}, in a suitable asymptotic limit.

The whole process of the  ID-state appearance in the wake of the obstacle, for
$u_m$ below $u_m*$, to its disappearance for $u_m$ above $u_m^*$, takes place in a very small interval  of values of the potential amplitude (Fig.~\ref{fscloseup}a-d). 
This  is a direct consequence of the ID-state instability, as we now show.
When $u_m$ is close to the critical amplitude $u_m^*$, the stationary solution $\phi(y)$ of Eq.~(\ref{eq2}) is close to the critical solution $\phi^*(y)$ for negative $y$ and for positive $y$ in the vicinity of the obstacle. Namely, on a length scale of order one behind the obstacle, one has $\vert \phi(y)-\phi^*(y)\vert\sim \vert u_m-u_m^*\vert$. This is also the magnitude of the 3 divergent modes in the vicinity of  the potential. Behind the obstacle, the distance between $ \phi(y)$ and $ \phi^*(y)$ grows exponentially and is dominated by the rate  $q_+$ of the fastest growing mode,
computed in Appendix \ref{appA}. 
 The front in the obstacle wake, which links the ID-state to one of the homogeneous stable states (Fig.~\ref{fscloseup}a-b), appears when  $\vert \phi(y)-\phi^*(y)\vert$ reaches  a value of order one. As a consequence, the distance $L$ of  the front from the obstacle is related to the difference $\vert u_m-u_m^*\vert$ between the potential amplitude and the critical one by,
\begin{equation}
\vert u_m-u_m^*\vert \exp(q_+ L) = O(1).
\label{asid}
\end{equation}
This explains the sharpness of the transition observed in the numerical simulations (Fig.~\ref{fscloseup}a-d) since in order to obtain a front at a distance $L$ from the obstacle, $u_m$ should be exponentially close in $L$ to $u_m^*$. Conversely, the distance of the front from the obstacle only grows logarithmically with the departure of $u_m$ from the critical potential amplitude, as
\begin{equation}
 L\sim (-1/q_+)\ln\vert u_m-u_m^*\vert.
 \label{lvsq}
\end{equation}
The measured distance $L$ is plotted  vs. $\ln\vert u_m-u_m^*\vert$  in Fig.~\ref{fscloseup}f,  for $u_m$ close to the critical potential and two values of $k_0$. The asymptotic slope, $-1/q_+$, is also drawn, using the spatial growth rates computed in Appendix \ref{appA}.  The quantitative agreement appears very good 
\footnote{In Fig.~\ref{fscloseup}, the values of $u_m$ are provided with  high precision. They correspond to values obtained within our particular numerical discretization scheme with the chosen spatial step $\Delta y=0.05$. The obtained high precision critical values $u_m^*$ , of course, approximate the actual ones for the continuous GGPE much less precisely, only to a few percent. We provide the value with high precision since they are required  to test Eq.~(\ref{asid}). Using these high precision values is meaningful since Eq.~(\ref{asid}) holds as well for the discretized equation. In addition, the value of $q_+$ is only different by a few percent between the discretized GGPE and the continuous one, so we have used the continuous value to check Eq.~(\ref{asid}) in Fig.~\ref{fscloseup}f.  
\label{hprecision}}.

\section{Analysis of the transition : slowly-varying obstacles}
 \begin{figure}[!h]
\begin{center}
\includegraphics[height=.8\textwidth]{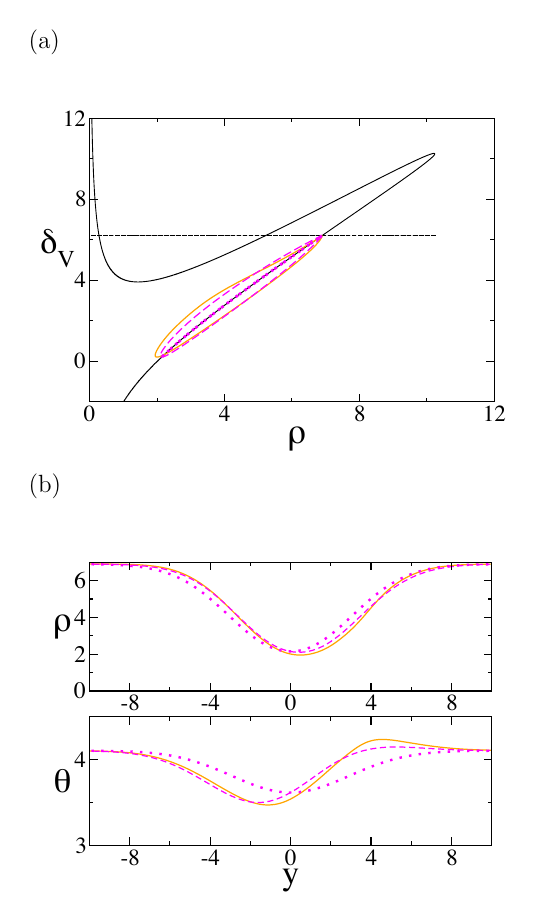}
\caption{Adiabatic approximation and corrections for the GGPE (Eq.~(\ref{eq2})). (a) Plot  in the ($\rho, \delta_V)$ diagram. (b) Plot of the density and the phase vs. $y$. The stationary numerical solution (orange solid line) of
Eq.~(\ref{eq2}) is plotted together with the adiabatic solution alone (dotted magenta; Eq.~(\ref{bvsrho}),(\ref{soladiaphase})) or with the  first-order corrections (dashed magenta; Eq.~(\ref{adiacor})). The parameters
are  $\sigma=4$,  $u_m=6$, $k_0=2.75$, $f=3.2, \delta_0=6.2$.)}

\label{adiaana.fig}
\end{center}
\end{figure}

In order to better understand these transitions and the role of bistability in the stationary flow metamorphosis, we  consider the
parameter regime suitable for theoretical analysis, provided by
an obstacle that varies on a long length scale, $\sigma\gg 1$ (Eq.~(\ref{gpot})).
 For a slowly varying obstacle, when the flow is accordingly slowly varying, the derivative terms in Eq.~(\ref{eq2}) can be treated perturbatively. At the lowest order, they can be entirely neglected and the ``adiabatic'' solution, $\phi_a(y)=\sqrt{\rho_a(y)} \exp[i\theta_a(y)]$, readily obtained.   The fluid density $\rho_a$,  is linked to the potential amplitude by Eq.~(\ref{fpe1}) with $\delta_0$ simply replaced by $\delta_V(y)$ (Eq.~(\ref{bdef})),
 which  takes into account the influence of the potential on the detuning. In this adiabatic approximation, solving the quadratic Eq.~(\ref{fpe1}) for $\delta_V(y)$ provides the implicit relation between the fluid density and the potential,
\begin{equation}
\delta_V(y)= \rho_a(y) - \sqrt{\frac{f^2}{\rho_a(y)}-1 }, \  \ 0\le \rho_a \le f^2.
\label{bvsrho}
\end{equation}
As for homogeneous solutions, the  solution phase is simply given as a function of the density
\begin{equation}
\theta_a(y)=\arctan\left(\frac{1}{\rho_a(y)-\delta_V(y)}\right)
\label{soladiaphase}
\end{equation}
The relation (\ref{bvsrho}) between the density $\rho$ and the ``detuning''  $\delta_V$  at fixed forcing amplitude  $f$, is plotted  in Fig.~\ref{adiaana.fig}a. It is equivalent but more convenient for our purpose than  Fig.~\ref{fullsim}a, which gives the density as a function of $f$ for fixed detuning.
A simple calculation shows that Eq.~(\ref{bvsrho}) determines the density as a unique function of $\delta_V$ when $f < f_c=(4/3)^{3/4}\simeq 1.2408$ while for $f>f_c$, there is a range
of $\delta_V$ values with multiple possible densities. In other words, bistability occurs for a range of $\delta_V$ values when $ f> f_c$, as illustrated in Fig.~\ref{adiaana.fig}a.

Let us now consider, a fluid injection in the HD-state, when the forcing is sufficiently strong for bistability to occur (i.e. $ f> f_c$). As the potential varies with the position $y$, $\delta_V(y)$ follows it according to Eq.~(\ref{bdef}). The density, as given by Eq.~(\ref{bvsrho}), moves along the HD branch in Fig.~\ref{adiaana.fig}a.
The adabatic solution is already
a close approximation of the flow obtained by numerically solving Eq.~(\ref{eq2}), for the Gaussian potential of Eq.~(\ref{gpot}) even with a rather  large amplitude ($u_m=6$) when
$\sigma=4$ (Fig.~\ref{adiaana.fig}b).

\subsection{The attractive potential case}
\begin{figure}[htbp]
\begin{center}
\includegraphics[height=1.1\textwidth]{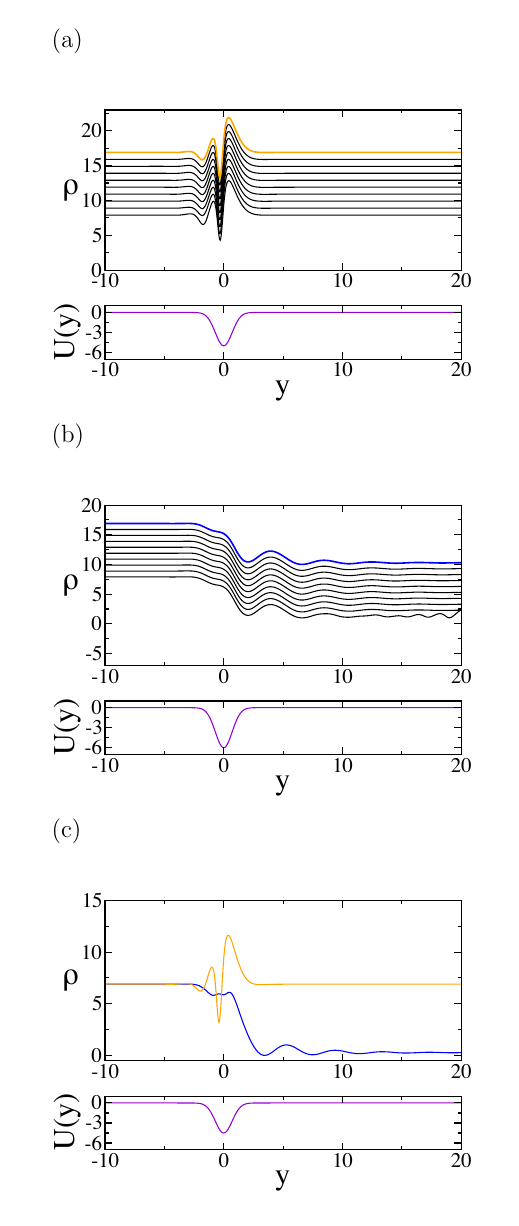}
\caption{Transition in the GGPE  for an {\em attractive} potential.
Simulations of (Eq.~\ref{eq2}) for  two localized potential amplitudes (a) $u_m=-5$ (b) $u_m=-6$. The density is shown at different times (solid black lines) with a time interval $\Delta\tau=5$ between the different curves.
The last simulation solution is plotted with a thicker line in (a) (orange solid line) and (b) (blue solid line). Successive curves have been shifted upward with time by 1 $\rho$-unit to show the stationarity of the flow. Note that the fluid beyond the obstacle is in the HD-state in (a) and in the LD-state in (b).
(c) The HD-state (orange solid line) and LD-state (blue solid line) solutions coexist in for $-5.1\le u_m \le -4.1$. They are displayed here for $u_m=-4.5$.
For each simulation, the dimensionless attractive Gaussian potential $U(y)$ is also shown (violet solid line), where  $U(y)=2 V(y)/(\hbar\gamma)$ (Eq.~(6)).
The other parameters are  $f=3.2, \delta_0=6.2, k_0=2.75, \sigma=1.$   }
\label{fS1}
\end{center}
\end{figure}

We first briefly describe the case of an {\em attractive} potential ($u_m<0$).  Eq.~(\ref{bvsrho}) predicts that the fluid density goes up the high density branch in Fig.~(\ref{adiaana.fig}, as $V(y)$ becomes more negative. The flow should undergo a transition if $\vert u_m\vert $ is large enough
for the top of the high density branch to be  reached,
since the branch cannot be followed  beyond its top. This transition is indeed seen in numerical simulations of Eq.~(\ref{eq2}) even away for the slowly-varying potential limit, as shown for $\sigma=1$ in
Fig.~\ref{fS1}.
As for repulsive potentials, for small $\vert u_m\vert $, the flow is stationary and in the HD-state in the wake of the obstacle (Fig.~\ref{fS1}a). There is a transition for a critical  amplitude $u_{m,1}$. When $\vert u_m\vert$ is larger than the critical amplitude $\vert u_{m,1}\vert $, the flow  in the wake of the obstacle is in the LD-state. The transition has however a different character than for repulsive potentials. The HD-state solution disappears at $u_{m,1}$ presumably by merging with an unstable solution in a classical saddle-node bifurcation. The LD-state solution exists and is stable below below  $u_{m,1}$. It disappears at
 $u_{m,2}$. Both solutions co-exist (Fig.~\ref{fS1}c)  when $u_m$ stands in between the two critical amplitudes, for $u_{m,1}<u_m<u_{m,2}<0$, which is therefore an interval of flow bistability.

\subsection{The repulsive potential case : a spatial rate-dependent tipping bifurcation}
How a transition can happen for a {\em repulsive} potential ($u_m>0$) is less obvious. The density of the adiabatic solution  (Eq.~(\ref{bvsrho})) follows the high density branch  towards low density before increasing again in the wake of the obstacle, as shown in Fig.~\ref{adiaana.fig}a.
This appears to be a smooth process for all potential strengths $u_m$. It is not clear why this would result in a transition of the flow profile at a critical amplitude $u_m$ and what this critical  amplitude would be.
 However, one can note that in the adiabatic approximation, all derivatives are absent and, as a consequence, the fluid velocity plays no role.
This suggests to go beyond the adiabatic approximation and treat perturbatively the derivatives terms.
Writing $\rho(y)= \rho_a(y) +\rho_1(y)+\cdots, \theta(y)= \theta_a(y) +\theta_1(y)+\cdots $, the first corrections to the adiabatic solution of Eq.~(\ref{bvsrho}), (\ref{soladiaphase}) are obtained after
a short calculation (see Appendix \ref{appB}) as,
\begin{equation}
\! \rho_1(y)=-\frac{ 2 k_0}{B'(\rho_a)} \frac{ d\rho_a}{dy} , \ \theta_1(y)=  \frac{ k_0}{B'(\rho_a)}
\frac{d}{dy}[\delta_V(y)-2 \theta_a]
\label{adiacor}
\end{equation}
where  $B'(\rho)$  denotes the derivative of $B(\rho)$ (Eq.~(\ref{fpe1})) with respect to $\rho$ .
These corrections are shown in Fig.~\ref{adiaana.fig}b and, as expected, they result in a closer agreement between the analytic approximations and the numerical  profiles.
More interestingly,
the corrected density profile  in the $(\rho, \delta)$ diagram provides a clue  to the origin of the instability (Fig.~\ref{adiaana.fig}a). One observes that the correction (\ref{adiacor}) produces a departure of the profile from the high density branch towards the middle unstable branch when the potential returns to 0, in the close downward wake of the obstacle. Eq.~(\ref{adiacor}) shows that this non-adiabatic effect grows with $k_0$ and,
 it also grows with the localized potential amplitude $u_m$. One can therefore guess, that, for sufficiently large $u_m$ or $k_0$, this leads the flow profile loop in
 Fig.~\ref{adiaana.fig}a, to reach the unstable density branch in the
$(\rho, \delta)$ diagram and leads to an instability. The global character of the bifurcation shows that it is invisible to linear (i.e.~Bogoliubov) excitations \cite{joly2021,stepanov2019,claude2022} around the steady flow. It cannot be located by a criterion that only involves them, like the Landau criterion for superfluids.
The bifurcation appears to be the analog in the spatial domain of ``rate-dependent tipping'' bifurcations \cite{ashwin2012,vanselow2019} in bistable systems which have
 become of interest in the context of climate change.
\begin{figure}[htb]
\begin{center}
\includegraphics[height=.8\textwidth]{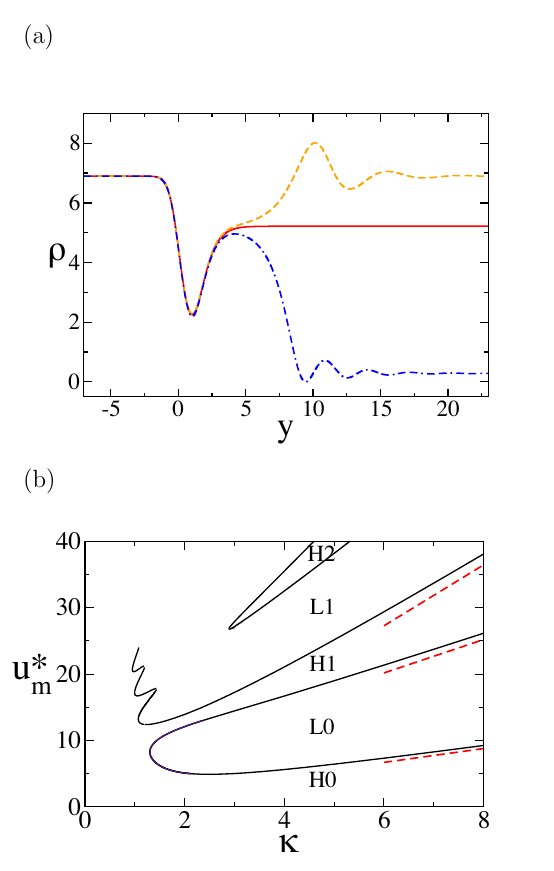}
\caption{Reduced asymptotic description. (a) Solutions of Eq.~(\ref{redreq}),(\ref{redpeq}) for $\kappa=2.75$. In the obstacle wake, the flow tends towards the HD-state for $u_m=4.9$ (dashed orange) and toward the LD-state for $u_m=5.0$ (dashed-dotted blue). The critical flow corresponds to $u_m^*\simeq 4.93$  (solid red) and tends to the ID-state. (b) Diagram of the transition lines in the $(\kappa, u_m)$ plane (solid black). The asymptotes for large $\kappa$ (Eq.~(\ref{vseries}), (\ref{slope}), (\ref{intero})) of the three lowest transition lines  are shown (dashed red). In the different parameter regions, it is indicated whether the flow in the wake of the obstacle tends toward the high (H) or the low (L) density states with the numbers corresponding to the additional $2\pi$ dephasing (Eq.~(\ref{vseries})) of the large $\kappa$ solutions in their downstream wake, as compared to the H0 and L0 ones. At finite $\kappa$, the boundary between different regions of the same type with different numbers (e.g. H0 and H1) is a line of solutions (not shown) with vanishing density at a point allowing the required phase jump.
 The zigzagging transition line at $\kappa\simeq 1$ has only been computed up to $u_m^*=24$. The other parameters are $f=3.2, \delta_0=6.2$.}
\label{redpb.fig}
\end{center}
\end{figure}

\subsection{Multiple transitions in a reduced asymptotic description.}
\label{asymp.sec}
 While suggestive, the above argument is not rigorous since the perturbative correction (\ref{adiacor}) cannot be trusted when it is not small.
In order to obtain a full reduced nonlinear description, a further asymptotic limit is needed, beyond that of a slowly varying potential (i.e. $\sigma\rightarrow\infty$). A simple mathematical one is obtained by increasing the flow velocity $k_0$ at the same time as the length scale of the potential is varied, i.e. taking the limit, $\sigma\rightarrow\infty,\, k_0\rightarrow\infty$ with a fixed ratio $\kappa =k_0/\sigma$. Determining the steady solution of Eq.~(\ref{eq2}) reduces in this limit to solving the simple system,
\begin{eqnarray}
\kappa \partial_z \rho& =& - 2 \rho - 2 f\sqrt{\rho} \sin(\theta)\label{redreq}\\
 \kappa \partial_z \theta &=& [\delta_V(z)-\rho] -f \cos(\theta)/\sqrt{\rho} \label{redpeq}
\end{eqnarray}
where $z=y/\sigma$.
Eq.~(\ref{redreq}), (\ref{redpeq}) simply give back  for $\kappa=0$ the adiabatic solution (\ref{bvsrho}),(\ref{soladiaphase}) and,  perturbatively for small $\kappa$, the correction (\ref{adiacor}). But, in the asymptotic limit considered, $\kappa$ can now take any value. The reduced system  (\ref{redreq}),(\ref{redpeq}), has only first-order derivatives in $z$.
Eq.~(\ref{redreq}), (\ref{redpeq}), have only one spatially-divergent mode from the unstable ID-state at $z=+\infty$. A simple shooting method thus determines the critical amplitude of the potential, $u_m^*$,  for which this divergence vanishes and  the solution tends at $z=+\infty$ toward the ID branch, as illustrated in Fig.~\ref{redpb.fig}a. For given driving parameters, multiple transitions are found by increasing the localized potential amplitude, as for the full GGPE. The fluid density in the wake of the obstacle is in the HD-state for low potential amplitudes. At  a first critical amplitude, it jumps to the LD-state, as described above. When the potential amplitude is further increased, a second transition is found, at which the fluid density jumps back to the HD-state. Further transitions are found  for still higher values of $u_m$. The loci of these  transitions are plotted in the $(k_0, u_m)$ parameter plane in Fig.~\ref{redpb.fig}b.

\begin{figure}
\begin{center}
\includegraphics[height=.8\textwidth]{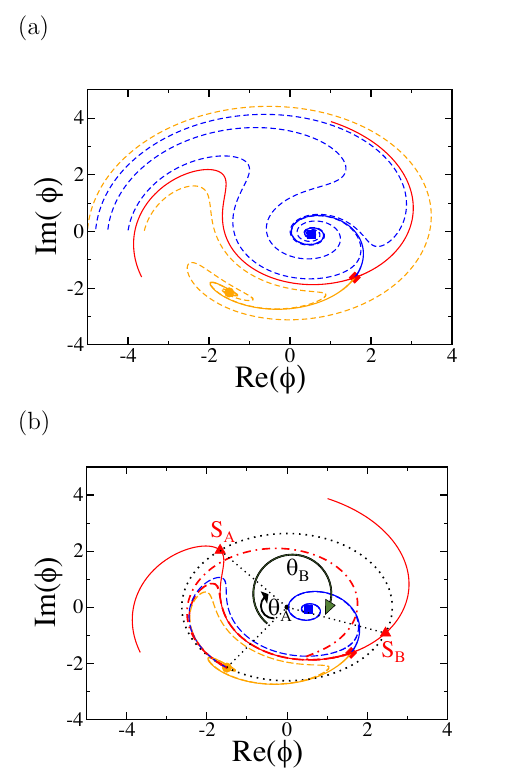}

\caption{
Phase plane analysis of the reduced asymptotic problem.
  (a) Phase plane of the homogeneous problem,  (Eq.~(\ref{redreq}), (\ref{redpeq}))  without the localized potential (i.e. $\delta_V(y)=\delta_0$) showing the three fixed points, HD (orange solid disk), ID (red solid diamond), LD (blue solid square)  together with the two entering separatrices (red solid lines) of the ID fixed point and the outgoing ones, ending on the HD point (orange solid line) or the LD point (blue solid line). Several trajectories are also shown ending either on the HD point (dashed orange line) of the LD-state (dashed blue line). (b)
Same diagram as (a) but showing two critical trajectories of the reduced problem (Eq.~(\ref{redreq}), (\ref{redpeq})) for $\kappa =6.0$ with the localized Gaussian potential of amplitude  $u_{m,1}^*\simeq 7.328$ (dashed red line) and
$u_{m,2}^*\simeq 21.34$ (dashed-dotted red line), corresponding to the first two transitions when $u_m$ is increased from 0. The critical trajectories start in the HD fixed point and they end on the ID fixed point.  Two other trajectories are shown which end either at the HD point, with  $u_m=6.5 < u_{m,1}^*$ (dashed orange line),
or at the LD point, with $u_m=8.0> u_{m,1}^*$ (dashed blue line).
It is also shown
a circle of radius equal to the modulus of the HD point (dotted black line) centered at the origin (solid black circle) as well as its two intersection points $S_1$ and $S_2$ (red triangles) with the entering separatrices of the ID point. The phase difference $\theta_1$ and $\theta_2$ between these intersection points and the HD point are indicated. They provide the asymptotic slopes
of the different transition branches (see Eq.~(\ref{vseries}) and the main text). The other parameters are  $f=3.2, \delta_0=6.2$.
}
\label{fS2}
\end{center}
\end{figure}

For the reduced  Eq.~(\ref{redreq}), (\ref{redpeq}), these multiple solutions and the asymptotics of the $u_m^*(k_0)$ branches can be analytically described  by considering
the large $\kappa$-limit. It is helpful to return to the complex function $\phi =\sqrt{\rho}\exp(i\theta)$  and analyze the dynamics of $\rho$ and $\theta$ in the complex
$\phi$-plane.
For large $\kappa$, the evolution  of $\rho$ and $\theta$  with $z$ is slow, except for the potential term $\delta_V(z)$ that evolves with $z$ on a scale of order 1. Apart from this fast action of the potential, the dynamics is governed by the phase plane of the problem without potential ($\delta_V(y)=\delta_0$). It is plotted in Fig.~\ref{fS2}a together with the 3 fixed points and a few trajectories. In the presence of the potential, a solution, that starts in the HD-state at $z=-\infty$, remains in the HD-state until it encounters the localized potential. The potential does not explicitly appear in Eq.~(\ref{redreq}) that governs the evolution of the density. The density change  produced by the potential is  mediated by the change of the phase $\theta$. It is  smaller than it by a factor $\kappa$, on  the length scale of order 1 where the potential has a significant amplitude. Therefore, at lowest order, the density does not change, on this length scale of order 1. On the contrary, Eq.~(\ref{redpeq}) shows that the phase $\theta$
rotates by an angle $\Delta\theta$,
\begin{equation}
\Delta \theta= -\frac{2}{\hbar \gamma\kappa}\int_{-\infty}^{+\infty} dz\, V(z) =-\sqrt{\pi}\,u_m/\kappa
\label{tturn}
\end{equation}
where the second equality holds for the Gaussian potential (\ref{gpot}). Since the density is conserved, the action of the potential is simply to displace $\phi$ on the circle of radius $r_H=\sqrt{\rho_H}$, where $\rho_H$ is the density of the HD-state, as shown in Fig.~\ref{fS2}b. For the solution to end up at $z=+\infty$ in the ID-state, the phase turn $\Delta \theta$ has to bring $\phi$ precisely, on one of the two entering separatrices of the  unstable ID-state, namely at their crossing points points $S_A$ or $S_B$, with the circle of radius $r_H$, as shown in Fig.~\ref{fS2}b. Therefore, $\Delta \theta$ should be equal to  $ \theta_{A,B}-2n\pi,\ n=0,1,\cdots$ where $\theta_{A,B}$ are the rotation angles corresponding to the displacement of the HD-state onto $S_A$ or $S_B$ (Fig.~\ref{fS2}b) .  The angle values with $n\ge 1$ correspond to the solution phase making full rotations before reaching one of the two separatrices. The double series of critical potential amplitudes for $\kappa\gg 1$, follows from Eq.~(\ref{tturn}),
\begin{equation}
u_m^* = -\kappa\, (\theta_{A,B} -2n\pi)/\sqrt{\pi} + O(1) \ \ n=0,1,2,\cdots
\label{vseries}
\end{equation}
For the parameter values of Fig.~\ref{redpb.fig} \& \ref{fS2} one has
$\theta_A\simeq -1.851, \theta_B\simeq -4.589 $.
Eq.~(\ref{vseries}) shows that asymptotically the critical potential amplitudes $u_m^*$ depend linearly on $\kappa=k_0/\sigma$. Eq.~(\ref{vseries}) gives the slopes of the asymptotic lines of critical potential amplitudes as a function of $\kappa$. In order to fully obtain the asymptotic lines, one also needs to compute the constant, next-order, term in the large $\kappa$- expansion of $u_m^*$, as derived in Appendix \ref{asymp.sec}. The obtained asymptotic lines for the three lowest branches, with the slopes given by Eq.~(\ref{vseries}) and the intercepts at the origin given by Eq.~(\ref{intero}), are displayed in  Fig.~\ref{redpb.fig}b together with the numerically obtained solutions. The two lowest transition branches merge at $\kappa\simeq 1.30$. The 4th and higher branches cross and recombine at intermediate $\kappa$ values, producing the bifurcation diagram  shown in Fig.~\ref{redpb.fig}b .

 Finally, one can note that Fig.~\ref{redpb.fig}b resembles Fig.~\ref{fS3}a,b for the full GGPE where $\sigma$ is not large. One difference is that the band H1 in Fig.~\ref{redpb.fig}b  terminates and does not exist at low  $\kappa$ values in Fig.~\ref{fS3}a,b, presumably due to the recombination of the 2nd (H0$\rightarrow$L1) and 3rd transition (H1$\rightarrow$L1). For higher values of $\kappa$, the transition lines in the full GGPE are close to that of the reduced model. 

\section{Conclusion}
In summary, the presence of an extended resonant drive and the  bistability that it creates, deeply change the transition of a condensate flowing past an obstacle.
The stationary flow profile undergoes a metamorphosis through the spatial analog of a rate-dependent tipping bifurcation  instead of  becoming time-dependent. The metamorphosis takes place in a very
small range of obstacle strengths (or velocities) due to the unstable nature of the wake at the transition.
Moreover, for given flow and pumping parameters, successive transitions exist at a discrete number of potential amplitudes.
We have shown that this can be
understood analytically in  a suitable asymptotic regime.  

It is worth emphasizing that the bifurcations of the flow profile that we have described,  are quite different  from usual textbook bifurcations. The steady state solution does not disappear by merging with an unstable solution, like in a saddle-node bifurcation, the bifurcation which, for instance,  gives rise to grey soliton emission in a one-dimensional condensate flow past and obstacle. The steady state above the bifurcation is also obviously different from the limit cycle in a  Hopf bifurcation which gives rise to vortex emission  in usual fluid flow or higher dimensional condensates.  Here, the  steady solution  persists through the bifurcation. It is stable above the bifurcation but its shape has abruptly changed. The stationary solution undergoes a metamorphosis at the bifurcation point, in a way that is only possible in an infinite dimensional system, namely by deforming at infinity.

The results  suggest a careful reexamination of the analogous flow transition in higher dimensions. We hope that they will also motivate experimental studies of the phenomenon. 
It will certainly be challenging to  experimentally resolve the details of the transitions and to witness the
appearance of the unstable state in the wake of the obstacle since this happens in a small neighborhood of the transition points. However, the transitions from a fluid in the HD-state to a fluid in the LD-state in the wake of the obstacle  and the steadiness of the flow both below and above  the transitions should be more easily observable.
Finally, we cannot help but wonder, whether the extended switches of the fluid density wake induced by a localized obstacle could provide useful applications in all-optical technology and devices \cite{ballarini2013}.

\begin{acknowledgments}We would like to thank A. Bramati, E. Giacobino, K. Guerrero and M. Jacquet for fruitful discussions. VH is also thankful to H Chat\'e, T Geisel  and the Heraeus foundation for their invitation to a workshop in Les Houches and the opportunity to learn about rate-dependent tipping bifurcations in other contexts.
\end{acknowledgments}

\appendix
\section{Stability of the constant density solutions}
\label{appA}
Here and in the following appendices, we find it convenient
 to analyze the GGPE (Eq.~(\ref{eq2})) by writing its solution $\phi$  as $\phi(y)= r(y)\exp[i\theta(y)]$, where the modulus $r(y)$ is the square root of the fluid density $\rho(y)$, $r(y)=\sqrt{\rho(y)}$. With these notations, the adimensionned  GGPE (Eq.~(\ref{eq2})) gives for the modulus $r$ and phase $\theta$,
\begin{eqnarray}
\partial_{\tau} r &= &- \frac{1}{2} [2\partial_y\theta\,\partial_y r+r\partial_{yy}\theta ]-k_0\partial_y r- r -f \sin(\theta) \label{req}\nonumber\\
r \partial_{\tau} \theta &=& \frac{1}{2} [\partial_{yy} r -r (\partial_y \theta)^2] -k_0 r\partial_y \theta  \label{phieq}\\
 & &\ \ \ \ \ \ \ \ \ \ \ \ \ \ \ \ \ \ \ \ \ \ \ \ + (\delta_V(y)-r^2)r -f \cos(\theta)\nonumber
\end{eqnarray}

We first   analyze the stability of the constant homogeneous solutions without potential ($\delta_V=\delta_0$).
 Their modulus $r_0$ and phase $\theta_0$ obviously obey
\begin{equation}
r_0 = -f  \sin(\theta_0), \, (\delta_0-r_0^2) r_0 = f \cos(\theta_0).
\label{fpe}
\end{equation}
Taking the square of each of these two equations and adding them,  gives back the previous Eq.~(\ref{fpe1}).

Linearization of the dynamical system (\ref{phieq}) around one such constant solution, $ r=r_0 +r_1, \, \theta=\theta_0 + \theta_1$,  shows that the first-order terms $r_1$ and $\theta_1$ obey,
\begin{eqnarray}
(\partial_{\tau}  + k_0\partial_y) r_1&= &- \frac{r_0}{2} \partial_{yy}\theta_1 - r_1 -f \cos(\theta_0) \theta_1 \label{lin1}\\
r _0(\partial_{\tau}   + k_0 \partial_y) \theta_1 &=& \frac{1}{2} \partial_{yy} r_1 +(\delta_0-3 r_0^2) r_1 +f \sin(\theta_0)\theta_1\nonumber \label{lin2}
\end{eqnarray}
Since the linear system (\ref{lin1}) is invariant by translation, we can look for the eigenvectors as Fourier modes, under the form
$r_1(y,t)=\overline{r}_1\,\exp(s t +i k y), \theta_1=\overline{\theta}_1 \exp(s t +i k y)$. The evolution of $(r_1, r_0 \theta_1) $ is governed by the  matrix $\mathbf{S}$ which has
$s+ i k k_0 $ as its eigenvalues, with
\begin{eqnarray}
\mathbf{S} &=&
\begin{pmatrix}
-1 &\ \ -f\cos(\theta_0)/r_0+ k^2/2\\
\delta_0 -3 r_0^2 -k^2/2 & \ \ +f \sin(\theta_0)/r_0
\end{pmatrix}
\nonumber\\
& =&
\begin{pmatrix}
-1 &\ \ r_0^2-\delta_0+ k^2/2\\
\delta_0 -3 r_0^2 -k^2/2 & \ \ -1
\end{pmatrix}
\label{smat}
\end{eqnarray}
The phase $\theta_0$  has been eliminated in the second equality with the help of the fixed point equations (\ref{fpe}).
The trace of $\mathbf{S}$  is negative, equal to $-2$. Therefore, the system is stable if and only if the determinant of $\mathbf{S}$ is positive,

\begin{eqnarray}
\mathrm{det} (\mathbf{S})&=& 1 -(r_0^2-\delta_0+ k^2/2)(\delta_0 -3 r_0^2 -k^2/2 )\nonumber\\
                                       &=& B'(r_0^2) +k^2 (2 r_0^2-\delta_0)+ k^4/4
                                       \label{dets}
\end{eqnarray}
where the function $B(\rho)$ in the last equality is defined by Eq.(\ref{fpe1}). The stability at long wavelengths ($k\ll 1)$) simply depends  on the sign of $B'(r_0^2)$ with $B'(r_0^2)<0$ leading to instability. That is, when there are multiple
 solutions, the branch of intermediate values of $r_0^2$  is unstable, the other ones are stable to homogeneous perturbations.

For Eq.~(\ref{lvsq}) and the counting argument of section \ref{numsim}, the values of
 the spatial growth rates $q=ik$ of  stationary perturbations (e.g. with $s=0$) are needed. They obey
 \begin{equation}
 B'(r_0^2)+2 q k_0+ (q k_0)^2 [1- \frac{1}{k_0^{2}} (2 r_0^2-\delta_0)] + \frac{(q k_0)^4}{4k_0^4}=0
 \label{sgr4}
\end{equation}
In order to determine the spatial growth rates about the LD, ID and HD states, one should first compute the coefficients of Eq.~(\ref{sgr4}), namely 
the densities of the LD, ID and HD states and the corresponding values of the $B$ function derivative. The three densities
are the roots of Eq.~(\ref{fpe1}). With our parameter choice of $f=3.2, \delta_0=6.2$, they are respectively equal to,
$\rho_{L}=r_{0,L}^2\simeq 0.2845,\ \rho_{I}=r_{0,I}^2\simeq 5.219,\ \rho_{H}=r_{0,H}^2\simeq 6.896$. The corresponding $B'(\rho)$ are
$B'(\rho_{L})\simeq 32.63, B'(\rho_{I})\simeq-8.276, B'(\rho_{H})\simeq 11.089 $.

For the ID-state, when  $k_0=2.75$, the four roots $q k_0$ of Eq.~(\ref{sgr4}) are found to be equal to $\{-5.82, 1.62- i\,11.1, 1.62+ i\, 11.1, 2.58\}$. Thus, as stated in the main text, there are 3 modes that exponentially grow with $y$. One is real positive and the two others are complex conjugate modes with a positive real part. The fastest spatially growing is the real mode
 $q_+= 2.58/k_0= 0.938$.
The corresponding slope $1/q_+= 1.07$ is used to plot the asymptotics of the intermediate state length as function of the departure from the critical potential amplitude (Eq.~(\ref{asid})) in Fig.~\ref{fscloseup}e (red dashed line). The situation is similar for $k_0=6.4$ with $k_0 q_+= 2.12$ for the fastest growing mode. The corresponding slope $1/q_+= 0.33$ is also shown in Fig.~\ref{fscloseup}e (red dashed-dotted line). We note that for large $k_0$, the case of interest for slowly varying potentials,  Eq.~(\ref{sgr4})  for $s=q k_0$ reduces to the 2nd order equation for the growth rate $s$ of an homogeneous perturbation of the ID-state. Namely, the fastest spatially growing perturbation simply corresponds to the advection of the unstable ID-state temporally growing mode.

For the HD-state, when  $k_0=2.75$, the four roots $q k_0$ are found to be equal to  $\{-5.26 - i\, 2.34, \, -5.25914 + i\, 2.34,\,
   5.26 - i\, 7.0 ,  5.26 + i\, 7.0 \}$. Therefore, there are two diverging modes  when $y$ tends towards $-\infty$ or $+\infty$, as stated in the main text.

\section{Expansion for a slowly-varying potential.}
\label{appB}
We provide here a derivation of the expressions for the adiabatic solution modulus (Eq.~(\ref{bvsrho})) and phase (Eq.~(\ref{soladiaphase})) and their first corrections (Eq.~(\ref{adiacor})).
We  suppose that the potential is slowly varying on an adimensioned length scale $\sigma\gg 1$ (as given by Eq.~(\ref{gpot})  for a Gaussian potential). We consider a stationary solution of the
GGPE as written in Eq.~(\ref{phieq}). We expands its modulus and phase as
$r(y)=r_a(y)+r_1(y)+\cdots, \ \theta(y)=\theta_a(y)+ \theta_1(y)+\cdots$ with $r_1$ and $\phi_1$ of order $1/\sigma$.
This gives
\begin{eqnarray}
r_a(y) +f \sin[\theta_a(y)] &=& 0, \label{rs}\\
\left(r_a(y)^2- \delta_V(y)\right) r_a(y) + f \cos[\theta_a(y)]&=&0 . \label{phis}
\end{eqnarray}
These are the same equations as those determining the constant solution with $\delta_0$ replaced by $\delta_V(y)$. The modulus $r_a(y)$  of the slowly varying solution corresponding to the stable HD branch  is given by Eq.~(\ref{bvsrho}).

 The first-corrections $r_1(y)$ and $\phi_1(y)$ obey,
 \begin{eqnarray}
r_1(y) +f \cos[\theta_a(y)]\, \theta_1(y)  &=& -k_0\partial_y r_a ,\label{r1}\nonumber\\
\left(3 r_a(y)^2 -\delta_V(y)\right) r_1(y) - f \sin[\theta_a(y)] \, \theta_1(y)&=&-k_0 r_a\partial_y \theta_a  .\label{phi1}\nonumber
\end{eqnarray}
This $2\times 2$ linear system is straightforwardly solved. The determinant of the matrix $\mathcal{L}$ on the l.h.s. is
\begin{eqnarray}
\det(\mathcal{L})&=& - f \sin[\theta_a] -f \cos[\theta_a(y)] \left[3 r_a(y)^2 -\delta_V(y)\right] \nonumber\\
&=& r_a(y) \left(1 +  \left[r_a(y)^2 -\delta_V(y)\right]\left[3 r_a(y)^2 -\delta_V(y)\right]\right)\nonumber\\
&=& r_A(y)
B'[r_a(y)^2], \label{det}
\end{eqnarray}
where we have used Eq.~(\ref{rs}) and (\ref{phis}) to express the phase in term of the modulus of the zeroth-order adiabatic solution and $B'(\rho) =dB/d\rho$ denotes the derivative of
the function  $B(\rho)$ (Eq.~(\ref{fpe1})). Similarly,  the Cramer's determinant for $r_1$ is
\begin{eqnarray}
\det(\mathcal{L}_r)&=&k_0f \left(\partial_y r_a \sin[ \theta_a(y)] + r_a \partial_y \theta_a \cos[\theta_a(y)]\right) \nonumber\\
 &=& k_0 f  \partial_y(r_a \sin[ \theta_a(y)] )=-k_0 \partial_y(r_a^2),\label{detr}
\end{eqnarray}
where  we have again used Eq.~(\ref{rs}) in the last equality.  The ratio the two determinants (\ref{detr}) and (\ref{det}) provide  the expression for the first correction to the modulus of the stationary slowly-vaying solution. Since the fluid density is the square of the modulus $r$, the first correction $\rho_1$ to the density is  $\rho_1=2 r_a r_1$. This finally gives Eq.~(\ref{adiacor}).

Similarly, one can compute the first-order phase correction. The Cramer's determinant for $\theta_1$ is,
\begin{eqnarray}
\det(\mathcal{L}_{\theta})&=& k_0 [3 r_a^2-\delta_V(y)] \partial_y r_a  - k_0 r_a \partial_y \theta_a \nonumber\\
  &=&  k_0 f \sin(\theta_a) \partial_y \theta_a+ k_0 r_a \partial_y \delta_V  - k_0 r_a \partial_y \theta_a \nonumber\\
  &=&  k_0 r_a \partial_a \delta_V  -2 k_0 r_a \partial_y \theta_a .
 \label{dett}
\end{eqnarray}
The ratio of the expressions (\ref{dett}) and (\ref{det}) for the two determinants gives  Eq.~(\ref{adiacor}) for $\theta_1$.

\section{Asymptotics in the slowly-varying potential and large flow velocity limit.}
We study  the asymptotics for large $\kappa$ of the reduced system described by Eq.~(\ref{redreq}), (\ref{redpeq}) with the potential $U(y)$
 \begin{equation}
U(z)=u_m\, u(z)
\end{equation}
For our choice of a Gaussian potential (Eq.~(\ref{gpot})), $u(z)$ is simply $\exp(-z^2)$.
We derive the leading estimation (Eq.~(\ref{vseries}))  of the critical potential amplitude as well  as the subleading constant term $u_m^{*,1}$ in its $1/\kappa$ expansion,
\begin{equation}
u_m^*=\kappa u_m^{*,0} +u_m^{*,1}+\cdots
\end{equation}

We seek the expansions of  the solution modulus $r(y)$ and phase $\theta(y)$, as well as the  potential amplitude, under the form
\begin{eqnarray}
r(z)&=&r_{H}+ \frac{1}{\kappa}\, r_1(z)+\cdots,\\
\theta(z)&=&\theta_0(z) + \frac{1}{\kappa}\, \theta_1(y)+\cdots,\\
u_m &=&\kappa u_m^0 +u_m^1+\cdots,
\end{eqnarray}
with the boundary condition $\theta(-\infty)= \theta_H$,
 where $r_{H}$ and $\theta_{H}$  are the phase and modulus of the  HD-state.

 At lowest order, Eq.~(\ref{redpeq}) gives
\begin{equation}
\partial_y \theta_0= -u_m^0 \, u(z),\, \mathrm{i.e.} \ \theta_0(z) =\theta_H -\int_{-\infty}^z\!\!\! dz'\,  u_m^0\, u(z'),
\label{as0}
\end{equation}
while $r$ remains constant equal to $r_{H}$. The critical potential amplitude $u_m^*$ is such that  the solution  asymptotically land on the ID fixed point. At the considered lowest order in $1/\kappa$, this requires that  the point $(r_{H}, \theta_0)$ should tend toward $S_A$ or $S_B$ (Fig.~\ref{fS2}),  the crossing points of the circle of radius $r_H$ with the two entering separatrices of the ID fixed point. Namely $\theta_0(z)$ should tend toward $\theta_S$, the angular coordinate of $S_A$ or $S_B$ when $z\rightarrow +\infty$. This  gives for the critical potential amplitude,
\begin{equation}
\int_{-\infty}^{+\infty}\!\!\! dz\,  u_m^{*,0}\, u(z)= \theta_H-\theta_S\ \mathrm{or}\  u_m^{*,0} =\frac{ \theta_H-\theta_S}{\int_{-\infty}^{+\infty}\!\!\! dz\,  u(z)}
\end{equation}
where $\theta_S$ should be equal to $\theta_H+\theta_A-2n \pi$ or $\theta_H+\theta_B-2n \pi$ in agreement with Eq.~(\ref{tturn}). Numerically, for the parameters of Fig.~\ref{redpb.fig} and Fig.~\ref{fS2}, one has  $\theta_A=-1.851,\, \theta_B=-4.462$ for the two separatrix points $S_A$ and $S_B$. This gives for the asymptotic slopes of the first three bifurcation branches (Fig.~\ref{redpb.fig}),
\begin{equation}
u_{m, 1st}^{*,0}=1.044,\  u_{m,2nd}^{*,0}=2.517,\  u_{m,3rd}^{*,0}= 4.589 .
\label{slope}
\end{equation}
where $u_{m,3rd}^{*,0}$ corresponds to the angle $\theta_A-2\pi$.

At the next order, the modulus $r_1$ and phase $\theta_1$ are given by,
\begin{eqnarray}
 r_1(z) &=& - \int_{-\infty}^z \!\!\! dz' \left\{r_H+ f \sin[\theta^*_0(z')]\right\} \label{r1y}\\
 \theta_1(z) &=&\int_{-\infty}^z \!\!\! dz' \left\{ \delta_0-r_H^2-f \cos[\theta^*_0(z')]/r_H -u_m^1 u(z') \right\}\nonumber \label{t1}
\end{eqnarray}
where the $*$-superscript on $\theta_0(y)$  is meant to denote that $\theta^*_0(y)$ is the solution of Eq.~(\ref{as0}) for $u_m^0$ equal to $u_m^{*,0}$.
The perturbation expansion used to obtain these expressions is valid as long as $r_1(z)/\kappa$ and $\theta_1(z)/\kappa$ are small. Namely, $z$ can be large but should be much smaller than $\kappa$. For $1\ll z\ll \kappa$, Eq.~(\ref{r1y}) and (\ref{t1}) show that $r_1(z)$ and $\theta_1(z)$ grow linearly with $z$ in a direction parallel to the separatrix at its crossing point with the circle of radius $r_H$,
\begin{equation}
r_1(z) \sim s_r z, \, \theta_1(z) \sim s_{\theta} z
\end{equation}
with
\begin{equation}
s_r=-r_H- f \sin(\theta_S), \ s_{\theta}= \delta_0-r_H^2 - f \cos(\theta_S)/r_H
\end{equation}
In order to reach the intermediate density point, the point $r(y), \theta(y)$ should belong to the separatrix when $y\gg1$. This gives the condition,
\begin{equation}
\lim_{y\rightarrow +\infty}[ s_{\theta} r_1(y)- s_r \theta_1(y)]=0
\end{equation}
 It determines $u_m^{*,1}$, the subleading term in the expansion of the critical potential amplitude, as
 \begin{eqnarray}
u_m^{*,1}&=&\frac{1}{\int_{-\infty}^{+\infty}\!\!\! dz\,  u(z)} \int_{-\infty}^{+\infty} \!\! dz \left ( \delta_0-r_H^2-\frac{f}{r_H} \cos[\theta_0(y)]\right.\nonumber\\
& & \ \ \ \ \ \ \ \ \ \ \ \ \ \ \ \ \ \ \ \ \ \left.+\frac{s_{\theta}}{s_r}\left \{r_H+f \sin[\theta_0(y)]\right \}\right )
\end{eqnarray}
For $f=3.2,\, \delta_0=6.2$, the modulus and phase of the HD-state are $r_H=2.626, \theta_H=-2.179$. With these values and the values of $\theta_A$ and $\theta_B$, one obtains for the subleading constants in the first three bifurcation branch asymptotics
\begin{equation}
u_{m, 1st}^{*,1}=0.428,\  u_{m,2nd}^{*,1}=5.074,\  u_{m,3rd}^{*,1}= -0.277 .
\label{intero}
\end{equation}

\label{asymp.sec}


\renewcommand\thefigure{S\arabic{figure}}
\setcounter{figure}{0}
\begin{figure}[htbp]
{\bf \Large Supplementary figure}
\begin{center}
\includegraphics[height=.78\textwidth]{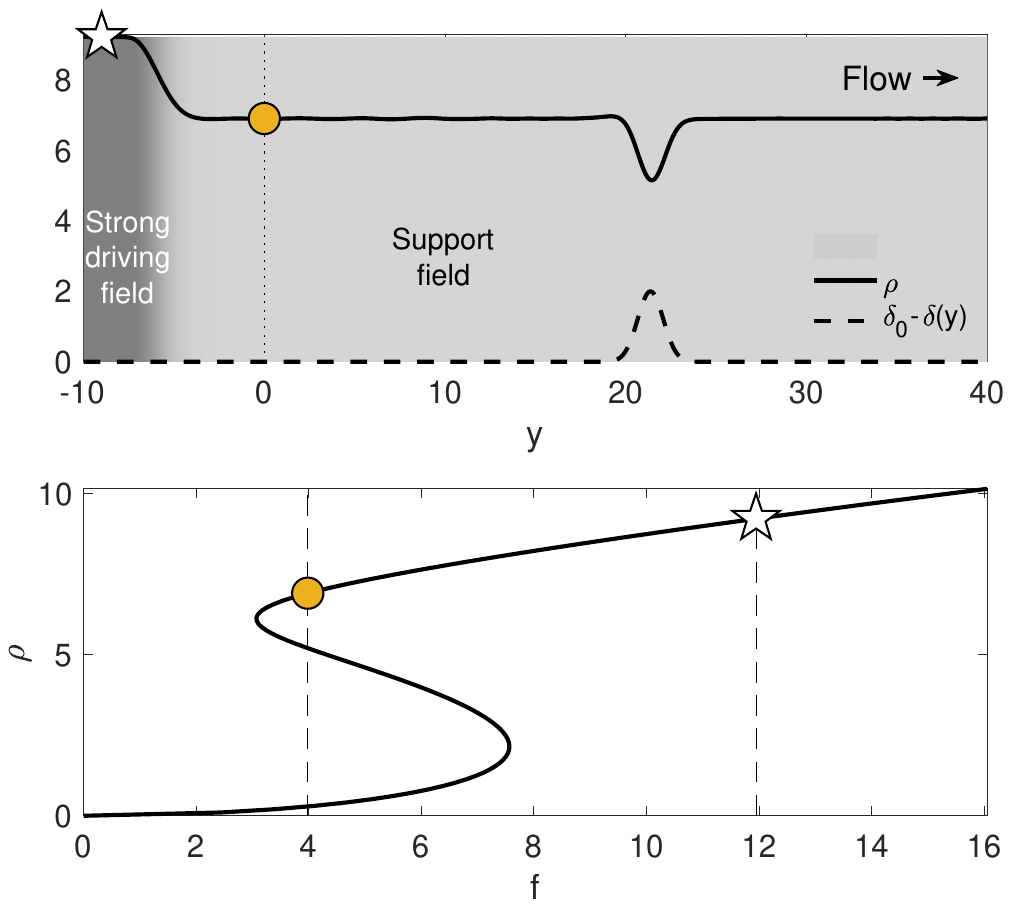}
\caption{Schematic description of the experimental setting. The top figure provides a sketch of the experiments (see e.g.  ref.[7] and [15] of the main text) that correspond to our theoretical description. The bottom figure shows where the fluid densities in the different regions are located in a bistability diagram (fluid density $\rho$ vs. amplitude of forcing $f$) similar to  Fig. 1a in the main text. On the far upstream side of the obstacle ($y<0$ region) a strong resonant driving field (dark grey shade) creates a high density of polaritons (white star in the top and bottom figure). Closer to the obstacle ($y>0$ region), the presence  of a a weaker ``support field''  makes the polariton density decrease and stay on the HD-state of a bistability window (solid orange circle) due to the presence of a weaker support field (light grey shade). In this second region, the fluid encounters a repulsive obstacle. In the simulations reported in the main text, only the fluid behavior in the light shaded region is simulated (light shaded region for $y>0$ starting at the dotted line). The upstream region with a strong drive is not present, it is simply taken into account by our upstream boundary condition with the fluid on the HD-state (orange circle) of the bistable region.
}
\label{fS0}
\end{center}
\end{figure}
\end{document}